\documentclass{aa/aa}  
\usepackage[utf8]{inputenc}
\usepackage{graphicx}
\graphicspath{{figures/}}
\usepackage{txfonts}
\usepackage{microtype}
\usepackage{hyperref}
\usepackage{siunitx}
\usepackage{subcaption}
\usepackage{xspace}

\newcommand{\src}{\object{Cen~X$-$3}\xspace}
\makeatletter
\renewcommand*\aa@pageof{, page \thepage{} of \pageref*{LastPage}}
\makeatother

\begin{document}

\title{Blind source separation for decomposing X-ray pulsar profiles}
   \subtitle{Introducing phase-correlated variability analysis (PCVA) with a case study of \src}

   \author{I. Saathoff      \inst{1}\inst{2}\inst{3} \and
           V. Doroshenko	\inst{1} \and
           A. Santangelo	\inst{1}}

   \institute{Institut f\"ur Astronomie und Astrophysik, Kepler Center for Astro and Particle Physics, Eberhard Karls Universit\"at, Sand 1, 72076 T\"ubingen, Germany\label{inst1}
   \and
   KTH Royal Institute of Technology, Department of Physics, SE-10691 Stockholm, Sweden\label{inst2}
   \and
   The Oskar Klein Centre for Cosmoparticle Physics, AlbaNova University Centre, SE-10691 Stockholm, Sweden\label{inst3}
   \\
       \email{saathoff@kth.se}}

   \date{Received September 15, 1996; accepted March 16, 1997}
 
   \abstract{
   Accretion-powered X-ray pulsars offer a unique opportunity to study physics under extreme conditions. To fully exploit this potential, the interrelated problems of modelling radiative transport and the dynamical structure of the accretion flow must, however, be solved. This task is challenging both from a theoretical and observational point of view and is further complicated by a lack of direct correspondence between the properties of emission emerging from the neutron star and observed far away from it. In general, a mixture of emission from both poles of the neutron star viewed from different angles is indeed observed at some or even all phases of the pulse cycle. It is essential, therefore, to reconstruct the contributions of each pole to the observed flux in order to test and refine models describing the formation of the spectra and pulse profiles of X-ray pulsars. In this paper we propose a novel data-driven approach to address this problem using the pulse-to-pulse variability in the observed flux, and demonstrate its application to \textit{RXTE} observations of the bright persistent X-ray pulsar \src. We then discuss the comparison of our results with previous work attempting to solve the same problem and how they can be qualitatively interpreted in the framework of a toy model describing emission from the poles of a neutron star.
   }

   \keywords{	Methods: data analysis -- 
   				X-rays: binaries --
   				pulsars: individual: \src --
   				Stars: neutron
               }

   \maketitle
   
\section{Introduction}\label{sec:introduction}
Accreting X-ray pulsars (XRPs) are neutron stars in a binary system fed by accretion from a non-degenerate (typically main sequence) donor star. The captured matter is funneled by a strong  (B $\sim 10^{12}$\,G) magnetic field to the poles on the surface neutron star where its gravitational energy is released, primarily in the form of X-rays. Neutron stars generally spin with the magnetic and rotational axes not necessarily aligned, so the emission observed by an external observer is often pulsed with the period corresponding to the rotation period of the neutron star. Readers can refer to \citet{White1983}, \citet{Nagase1989}, \citet{Bildsten1997}, \citet{Caballero2012}, and \citet{Mushtukov2022} for reviews on accreting XRPs.
\begin{figure*}[htbp]
  \centering
  \resizebox{\hsize}{!}{\includegraphics{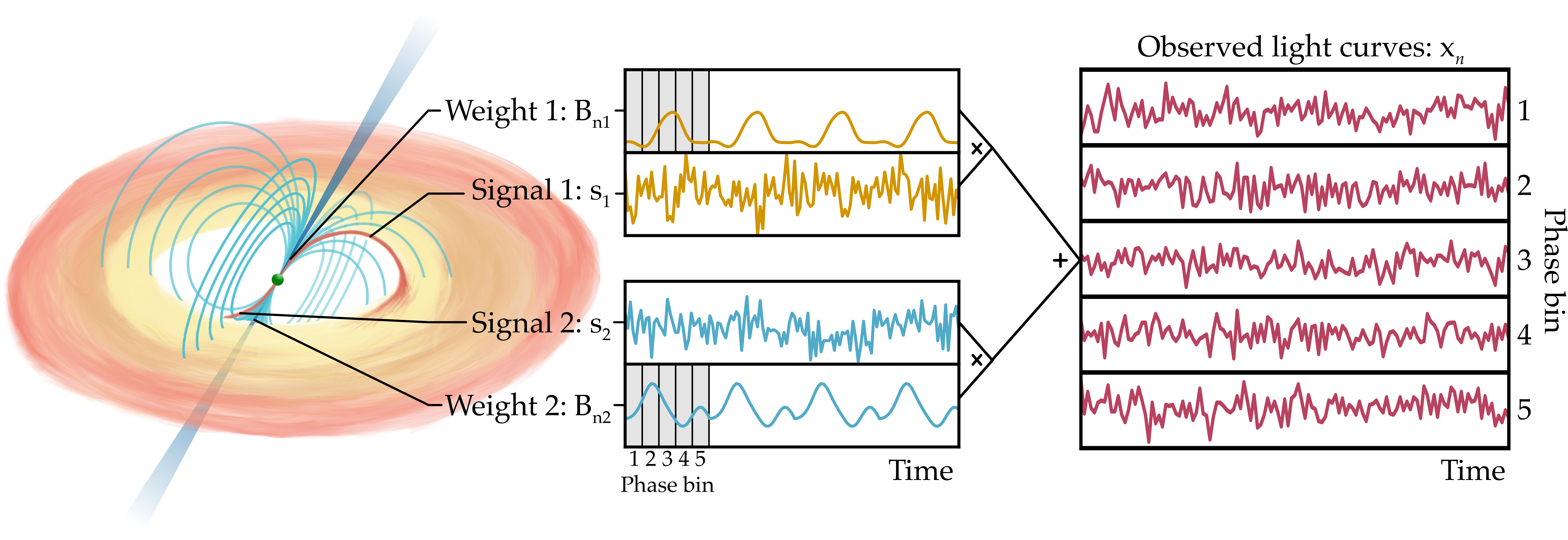}}
  \caption{Sketch of an accreting X-ray pulsar and the components that make up its observed light curve as a function of phase. At the distance of the inner accretion disk, the magnetic dipole field dominates and matter flows along the magnetic field lines onto the magnetic poles of the neutron star where it releases X-rays. We consider the accretion rates onto each of the poles to be our signals ($s_1$ and $s_2$) since fluctuations of the accretion rate are roughly translated into fluctuations of the observed X-ray flux; and the beam intensity at a given pulse phase is the single-pole pulse profile, also called the weights ($B_{n1}$ and $B_{n2}$ for $n$ phase bins). In the figure, the single-pole pulse profiles are shown in the grey box along with five phase bins as an example. The profiles were then repeated for the rest of the observation for clarity. These weights cause the X-ray flux at a given pulse phase to be systematically offset and the flux in two different phase bins appears correlated if the radiation at that phase originates from the same pole, but less correlated if both poles contribute. The observed phase-resolved light curves ($x_n$) are then the result of a mix of the emission from both poles.}
  \label{fig:sketch}
\end{figure*}
The regular variability pattern of the flux with the spin phase, called the pulse profile, is obtained by averaging these changes over several rotation periods. While the basics of pulse profile formation were understood early on (for a brief review, see \citealt{Fuerst2019}), the observation of complex profiles that vary with energy and luminosity is a challenge to any fully consistent theoretical interpretation.

To fully appreciate the complexity of the problem, it is necessary to understand the various parameters that contribute to the formation of a pulse profile. The magnetic field configuration of the neutron star, for example, plays a crucial role in determining the profile's structure. As matter couples to the magnetic field lines at the inner accretion disk, it encounters specific parts of the neutron star's surface, changing the shape and position of the footprint of the accretion flow on the surface. A displacement of the dipole, as well as the presence of higher multi-pole components, can also alter the emission region geometry \citep{Kraus1995, Romanova2004, Long2007}. In addition to the magnetic field configuration, the structure and physical parameters of the emitting region are largely determined by the feedback of the X-rays themselves, which in turn is influenced by the accretion rate \citep{Basko1976, Meszaros1984, Mushtukov2022}.  At low accretion rates and corresponding low luminosities ($L_X \leq 10^{37}$\,erg\,s$^{-1}$), an emission region appears on the neutron star's surface, and X-rays are typically emitted upwards in a pencil beam, parallel to the magnetic field. The shape of this emission region, whether a crescent- or ring-shaped emitting region \citep{Lamb2009}, and the formation of an accretion mound \citep{Mukherjee2012} further influences the pulse profile's shape. On the other hand, at high accretion rates and high luminosities ($L_X \geq 10^{37}$\,erg\,s$^{-1}$), a shock forms above the neutron star's surface \citep{Basko1976} with plasma slowly sinking below it. In this case, X-rays are more likely to be emitted towards the sides in a fan beam, perpendicular to the magnetic field. 
It is important to recognise that the emission represented as a pencil or fan beam is a highly simplified approximation, and the actual emission is much more complex.
Whether the resulting accretion column is more like a filled cylinder or a thin-walled hollow funnel could depend on the size of the threading region in the inner accretion disk \citep{Basko1976}. An uneven contact between the accretion disk and magnetosphere could break up the structure into segments or cause deviations from a spherically symmetric flow. Additionally, instabilities near the polar cap could cause matter to reach the surface as pancakes or spaghetti. The emerging X-rays that are emitted near the surface of the neutron star are reflected off the surface or bent by the strong gravitational field. These processes can affect the intensity of the observed X-rays and change its apparent origin \citep{Riffert1988, Kraus2001, Poutanen2006, Poutanen2013, Mushtukov2022}. Finally, the geometric configuration of the system plays an important role. The angle between the neutron star's spin axis and magnetic axis, as well as angles between the spin, binary orbital angular momentum, and line of sight can influence what is observable \citep{Meszaros1988}.
To constrain the geometry of the emission region and the physical processes involved, an essential first step is to understand what we actually observe, that is, understand the contributions of the individual emitting regions to the observed pulse profiles. This can be, however, a challenging task as the emission from both poles may contribute to the signal observed at a given pulse phase due to the reasons mentioned above, including beam patterns, column geometries, and the gravitational bending and beaming of X-rays \citep{Mushtukov2022}.

Several attempts to decompose pulse profiles into its components and understand the geometry of the emission regions have been done in the past. The most straightforward approach is to model the contributions of individual emitting regions based on some basic assumptions regarding the intrinsic beam pattern of each of the poles and using relativistic ray-tracing to predict the observed flux for various pulsar geometries (i.e. different orientations of the spin and magnetic axes relative to the line of sight). For example, \citet{Iwakiri2019} assumed that the intrinsic beam pattern is a combination of fan and pencil beams, while \citet{Sokolova-Lapa2021} described the intrinsic beam pattern as two separate components emitted from the walls and the top of the accretion column. However, this approach relies on the adequacy of the assumed intrinsic beam pattern and theoretical models used to calculate, which are hard to assess independently.

An alternative approach was pursued by \citet{Kraus1995} who proposed a data-driven pulse profile decomposition method based on assumption of intrinsic symmetry of flux emerging from each of the poles. Their method has been applied to several sources (\src \citep{Kraus1996}, Her~X-1 \citep{Blum2000}, EXO~2030+375 \citep{Sasaki2010}, 1A~0535+26 \citep{Caballero2011, 2023ApJ...945..138H}, 4U~0115+63 and V~0332+53 \citep{Sasaki2012}), and has enabled the reconstruction of the geometry and intrinsic beam patterns. The pulse profile of \src has been studied by \citet{Kraus1996} and their decomposition result, shown in \autoref{fig:kraus}, yields two symmetric but rather dissimilar and complex components. We note that recently \cite{2022NatAs...6.1433D} and \cite{2022ApJ...941L..14T} compared the phase-dependent polarisation degree to the results of \citet{Kraus1996} and found good agreement between the polarisation degree and the relative contribution of one of the poles to the total flux at given pulse phase. We note, however, that the reason for this is unclear, as mixing of emission from two poles does not necessarily lead to a reduction in polarisation, especially if the observed polarisation properties are altered by propagation through the magnetosphere, as suggested by the good agreement of the observed polarisation angle dependence on pulse phase with a simple rotating vector model \citep{1969ApL.....3..225R, 2022NatAs...6.1433D, 2022ApJ...941L..14T}. Moreover, recent observation of transient XRP RX~J0440.9+4431 showed that both polarisation degree and angle can be affected by additional polarised components arising, for instance through scattering in stellar wind of the companion \citep{2023A&A...677A..57D}. A similar scenario (i.e. scattering off the accretion disk or in the stellar wind) is also likely to be relevant for Cen~X$-$3 and Her~X$-$1, where several observational features point to the presence of a strong disk wind \citep{2011ApJ...737...79N, 2011ApJ...731L...7M, 2022ApJ...927..143W, 2023arXiv230405412N, 2023ApJ...959...51K}. 
We believe, therefore that polarimetric observations should not be viewed as direct confirmation of the robustness of \cite{Kraus1996} method and other opportunities shall be explored.

This is especially true as one can point to several potential issues associated with the \cite{Kraus1995, Kraus1996} approach.
First, the basic assumption regarding the symmetry of the individual pole contribution in phase space is indeed not sufficient to unambiguously identify the correct decomposition. In fact, an infinite number of potential solutions are possible, corresponding to arbitrary choices of symmetry points in phase space. The selection of the final solution is based on a set of ad hoc selection criteria (for a summary of these criteria, refer to \citealt{Sasaki2010}), which could potentially lead to incorrect results. 
Furthermore, the method of \citet{Kraus1995} is fundamentally based on the assumption that the emission originates from the magnetic poles with identical emission characteristics and a symmetric beam pattern. The observed pulse profiles of X-ray pulsars are asymmetric and thus can only be approximated by the sum of the two symmetric components if they are offset from each other. \cite{Kraus1995} attributed such an offset to a slight distortion of the magnetic dipole field, resulting in the axis connecting the two poles not passing through the centre of the neutron star. The main issue here is that the basic assumption of intrinsic symmetry is actually hard to justify, especially for extended emission regions such as accretion columns \citep{Basko1976} or arc-like extended polar caps \citep{Basko1976,Romanova2004}.
In fact, even for small emission regions on the surface, the angular distribution of the resulting flux may not be symmetric. Indeed, \citet{Kaminker1983} and \citet{Burnard1988} show that this is only true if the magnetic field is perpendicular to the atmospheric surface, which is not a given. For instance, this is likely not the case if the surface field has strong multipole components or if there is an extended accretion mound or column. For instance, \citet{Mukherjee2012} shows that an extended accretion mound could cause the magnetic field lines to bend in order to support the pressure of the confined matter, resulting in a distortion of the magnetic field relative to the normal of the plasma. The situation for accretion columns is even less clear, but the photosphere of the neutron star can also be expected to be at some angle relative to the field lines \citep{1988SvAL...14..390L}.

It is important to develop an independent method whose results can be compared with those of \citet{Kraus1995}.
Here we describe a novel data-driven technique called phase correlated variability analysis (PCVA) to address this problem. The main idea behind this method is to exploit the observed variability of the flux from each pole with no further assumptions regarding the symmetry of the contribution from each pole. 
In \autoref{sec:method} we provide a detailed description of the PCVA methodology, test its reliability, and discuss the requirements for observational data to ensure that the reconstruction is robust. In \autoref{sec:application}, we apply the method to \src and present the results of the PCVA analysis. These are then discussed in the context of previously published results and predictions of a toy model that we develop to qualitatively describe the emission from emission regions on the surface of an X-ray pulsar, where the angular distribution of the X-rays emerging from the poles is not necessarily symmetric. We conclude with a summary in \autoref{sec:summary}.

\section{Phase correlated variability analysis (PCVA)}\label{sec:method}
\subsection{Description of the PCVA method}

\begin{figure*}[htbp]
    \centering
    \resizebox{\hsize}{!}{\includegraphics{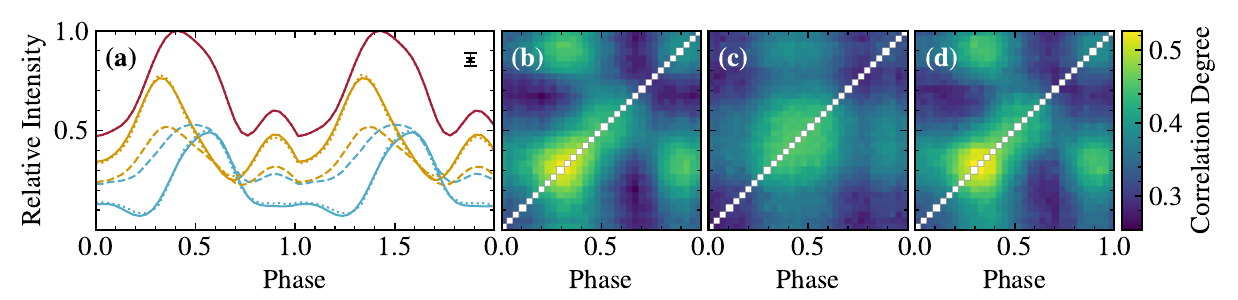}}
    \caption{Results of the \textit{NMF} decomposition and rescaling of an example simulated dataset. In panel (a), the simulated contributions of the two poles of an XRP are shown as blue and yellow solid lines. The sum of these components is the `observed' pulse profile (red solid line). The resulting contributions of the \textit{NMF} decomposition, shown as blue and yellow dashed lines, are close to the input contributions (solid), only the relative scaling is not recovered fully. After optimisation using the correlation properties, the recovered signals are closer to the input (blue and yellow dotted lines). The estimated minimum and maximum uncertainties are shown in the upper right-hand corner. Two pulse phases are shown for clarity.
    Panel (b) shows the correlation matrix of the original simulated data. Panel (c) shows that when simulating data based on the two components of the \textit{NMF} decomposition, the correlation matrix appears to be less correlated than the original (b). Panel (d) shows the correlation matrix of a simulation based on the optimisation of scaling of the components. This results in a better representation compared to the input matrix (b).} 
    \label{fig:synthetic_dataset_pcva}
\end{figure*}

\begin{figure*}[htbp]
	 \resizebox{\hsize}{!}{\includegraphics{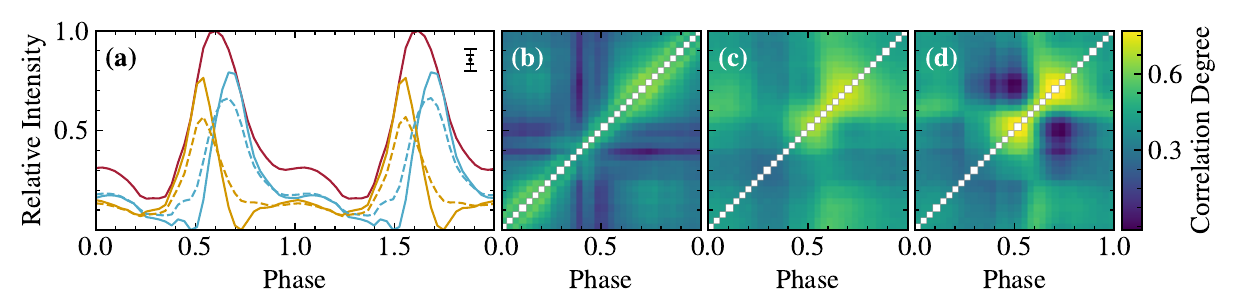}}
	 \caption{Results of the \textit{NMF} decomposition and rescaling on the basis of a real \textit{RXTE} observational dataset of \src. Panel (a) shows the pulse profile of \src as the red line and the average PCVA decomposition in blue and yellow. Two pulse phases are shown for clarity. The estimated minimum and maximum uncertainties are shown in the upper right hand corner of the panel. The two components are qualitatively similar when one of the components is reflected and shifted in phase. The angles between the minima and maxima of each component are $67.5^\circ$ and between the two components are $90^\circ$ and $45^\circ$. An example of the components prior to the rescaling are shown as dashed lines. Panel (b) shows the correlation matrix corresponding to the \textit{RXTE} observation of \src. The presence of distinct structures in the correlation matrix suggests that the phase-resolved light curves exhibit the correlation properties described in the text. Panels (c) and (d) show the correlation matrices of the phase-resolved light curves based on the PCVA decomposition before and after rescaling of the components, respectively.
    }
	 \label{fig:results}
\end{figure*}

Matter from the companion star of XRPs can either fall roughly spherically onto the surface of the neutron star, as in the case of direct wind accretion, or through an accretion disk. In either case, the accretion flow is eventually disrupted by the pulsar's magnetosphere and funneled to the surface, but the stochastic fluctuations in the accretion flow persist and are manifested as variability in the observed X-ray emission \citep{Lyubarskii1997, Revnivtsev2009}. This phenomenon is typical of accreting systems and occurs over a wide range of timescales \citep{Doroshenko2020, Moenkkoenen2022} likely defined by the characteristic dynamo timescale in the accretion disk \citep{Mushtukov2019}. Due to these fluctuations, which are expected to be different between the two poles, each individual pulse has a different shape \citep{Tsygankov2007, Klochkov2011}, which is the reason why averaging over many cycles is required to obtain the average pulse profiles of accreting pulsars.

Importantly, the variability occurs over a wide range of frequencies, including the spin frequency of the pulsar, and is expected to be largely independent for the two poles of an inclined rotator. As illustrated in \autoref{fig:sketch}, the accretion flow to each of the two poles is indeed expected to originate from opposite parts of an accretion disk that is close to co-rotation with the neutron star, and thus any variability on timescales close to or below the Keplerian timescale can be expected to be independent for the two poles. 
This can be used to disentangle the contribution of the two poles. Since the instantaneous accretion rate changes at each of the poles are independent, one might expect that on a time scale corresponding to the spin frequency of the neutron star (i.e. from one pulse to the next), parts of the pulse profiles dominated by accretion to a given pole change in concert, while parts dominated by the contribution from the other pole do not follow the same trend. One can therefore expect that pulse-to-pulse light curves extracted from individual phase bins are more or less correlated, and the degree of correlation can be used to constrain the contribution of each pole to individual phase bins.
We note that in principle a fraction of the photons could lose their phase information, for example due to scattering in the wind surrounding the neutron star. This would affect the observed degree of correlation and thus bias the method. However, the fraction of scattered flux is not expected to be too large and would mostly affect the absolute scaling of the recovered components, so it can be disregarded in a first order approximation.


The assumption of partial independence can be verified by observation. The degree of correlation between light curves at different phase intervals can be illustrated in a correlation matrix, an example of which is shown in panel (b) of \autoref{fig:results}. Since the correlation coefficient is not always one or zero, this indicates that the signals are not completely dependent on each other. Instead, the degree of correlation varies throughout the rotation period, leading to the conclusion that the assumption of partial independence is valid.
The correlation properties can be used to disentangle the flux from each pole. This involves separating two signals, namely the flux from each pole as a function of the pulse phase.
This task can be formulated as the well-known blind signal or source separation (BSS) problem, which aims at reconstructing some initial signals from mixed observations.\footnote{The often given example of a BSS problem is the cocktail party problem, where a guest is attempting to listen to one conversation amidst noisy surroundings. This is also known as selective hearing, and while humans can do this relatively easy, it is a difficult problem for machines to solve.} BSS can be found in many areas, such as telecommunications and medical applications. In medical applications, for example, it can be used to separate interfering muscle activity or other bodily functions from the brain activity being studied. In astrophysics, BSS methods are increasingly used for data analysis. Non-negative matrix factorisation (NMF), one of the methods for solving the BSS problem, has been used to refine the calibration of the Keck observatory spectrograph \citep{Horstman2022}. NMF has also been used to analyse images of exoplanetary systems \citep{Ren2018}, and in solar physics BSS has been used to study coronal temperature maps of the Sun \citep{Dudok2013}. These examples show that BSS and its methods are generally well established.

However, it needs to be shown that BSS is also suitable for the particular problem of pulse profile decomposition considered here. One of the problems here is that we are not actually observing flux variations, but rather count rate variations in a detector located far from the source. Indeed, even with large area detectors such as the Rossi X-ray Timing Explorer (\textit{RXTE}) \citep{Bradt1993} or the enhanced X-ray Timing and Polarimetry mission (\textit{eXTP}) \citep{2019SCPMA..6229502Z}, the total number of X-ray photons detected within a relatively small time interval corresponding to a phase of the pulse cycle may be dominated by Poisson statistics, and at low count rates the correlation properties mentioned above may be lost. This defines the limits of the applicability of correlation analysis methods, which will be discussed below. 
The limited counting statistics or, equivalently, the relatively high noise common to astronomical observations may reduce the effectiveness of some of the available BSS algorithms. Furthermore, additional steps specific to the problem of reconstructing the flux contributions of each of the two XRP poles to the observed pulse profile are required, so in this particular case we give a more specific definition of the BSS problem, namely PCVA. The core of the problem, and thus the goal of BSS and in our case PCVA, is to identify different source signals from mixed signals with very little knowledge of the source signals and the mixing process.

Much of the terminology we use remains the same as in classical BSS. In particular, the $m$ source signals can be represented as a vector
$$\begin{gathered}
    s(t) = [s_1(t), \dots, s_m(t) ]^\intercal , \\
\end{gathered}$$
where $\top$ is the transposition.
In our case, we assume that we have two source signals corresponding to a variable accretion rate at each of the poles traced by the observed X-ray flux variability, so $m=2$.\footnote{In principle, $m$ can be greater than 2, but we assume that the accretion rate is partially independent because it starts at the inner accretion disk -- where the magnetic dipole dominates. Thus, $m=2$ and any other choice of $m$ must be carefully checked to see if statistical independence is still expected.} However, the observed flux is not directly related to one of the two signals, but represents a mixture of them. Mathematically, the observed flux can thus be described as a mixture of the two signals with some mixing operator
$$\begin{gathered}
    x(t) = [x_1(t), \dots, x_n(t) ]^\intercal ,
\end{gathered}$$
where each mixed signal is given by $x = As(t)$ and $A = [a_{ij}] \in \mathbb{R}^{n \times m}$ is the mixing operator, which is a matrix of mixing coefficients corresponding to multiple observations (it is not possible to disentangle two signals using less than two observations). In our case, the mixed signals correspond to the observed count-rates $x(t)$ in $n$ phase bins, that is, the observed pulse-to-pulse light curves for each bin. The number of the `observations' is then equal to the number of phase bins used to construct pulse profile, and the mixing operator $A$ contains coefficients defining the contribution of each signal to a given phase bin, that is, it represents the decomposition of the observed pulse profile into two components corresponding to underlying accretion rate fluctuations $s(t)$. We note that neither $s(t)$ nor $A$ are actually known, and their estimation is at the heart of the BSS problem. In fact, in the classical BSS formulation, only the signals $s(t)$ themselves are of interest, whereas we are more interested in estimating the mixing operator, which is another reason why we use the PCVA term.

The goal of the PCVA is thus to recover the mixing operator (and the source signals) by finding a matrix $B = [b_{ij}] \in \mathbb{R}^{n\times m}$. This matrix $B$ is an approximation of the mixing matrix $A$ and is therefore not in general exactly equal to it. The source signals then correspond approximately to
$$\begin{gathered}
    y(t) = [y_1(t), \dots, y_m(t) ]^\intercal,
\end{gathered}$$
where $y = Bx(t)$. 
The source signals $s_{1,2}$ thus correspond to $y_{1,2}$ and represent the observed emission from the poles. With our method, we recover the source signals as well as the mixing operator $B$, which represents the beam intensity at a given phase.

Since BSS is generally an underdetermined problem, any method for solving it introduces some constraints or assumptions.
However, the only common assumption of all BSS algorithms is that the signals are at least somewhat independent, and that the solution sought (for both the signals and the mixing operator) corresponds to maximally independent signals. In our case, the accretion rate at two poles is expected to be correlated to some extent on timescales longer than the spin period, so the solution will be biased towards the pulse-to-pulse variability, which is independent. This means that the absolute scaling of the mixing operator is not expected to be accurately recovered, and additional steps, also considered as part of the PCVA, may be required to do so, as we discuss below.

Another difference compared to general BSS is that the observed flux (counts or count rates), and thus the components we want to retrieve, are always positive. This non-negativity is an important constraint on the solution, and one method that applies it is the non-negative matrix factorisation (NMF), also known as NNMF \citealt{Lee1999}). Therefore, we implement the NMF method in the PCVA. In Python, it is implemented in the Scikit-learn package \citep{Pedregosa2011}\footnote{\url{https://scikit-learn.org/stable/modules/generated/sklearn.decomposition.NMF.html\#sklearn.decomposition.NMF}}. We expect it to return results that are physical and easy to interpret. However, we have investigated several other algorithms, namely principal component analysis \citep{pearson_karl_1901_1430636}, independent component analysis (ICA)\citep{Cardoso1998MultidimensionalIC, HYVARINEN2000411}, and preconditioned ICA for real data \citep{ablin2017faster, Ablin_2018}.
When applying these algorithms to simulated data, we found that they all produce comparable results for high counting statistics. However, the NMF recovery was more stable and therefore has smaller uncertainties, so it is the natural choice for this reason as well. Details on the performance of different algorithms in different regimes will be published elsewhere and are not relevant for the current work. Last but not least, it is important to emphasise that the correlation matrices mentioned above are not actually used by NMF and are only plotted for illustration and visualisation purposes. The actual reconstruction is done directly using the pulse-to-pulse light curves corresponding to the observed flux in a given phase bin in a given pulse cycle.

\subsection{Verification of the method using simulations}
\label{sec:verification}
To demonstrate that BSS can indeed be used to reliably decompose pulse profiles in XRPs, we first verified the method using a series of simulations, where the analysis is performed on simulated light curves that contain information about both the underlying accretion rate fluctuations and the mixing defined by the mixing operator that characterises the beam intensity contributions of the two poles, also referred to as single-pole pulse profiles. It is these profiles that we aim to recover when applying the method to real data, but for simulations this is obviously a known input that can be compared with the reconstruction results to verify the robustness of the method. The simulation and verification steps are explained below.

\begin{table}
\caption{Parameters of the simulated light curves based on red-noise-type power spectral density spectra. The values are typical for accreting XRPs \citep{Fuerst2019, Moenkkoenen2022}.}
\label{tab:simulated lc params}
\centering
\begin{tabular}{lc}
	\hline\hline
	Parameter & Value \\ 
	\hline
	Exposure & 10\,ks \\
	Average count rate & 100\,cts\,s$^{-1}$ \\
	Spin period & 1\,s \\
	PSD normalisation & 1 \\ 
	Power law index 1 & 1 \\
	Power law index 2 & 2 \\
	Break frequency & 1\,Hz\\
	\hline
\end{tabular}	
\end{table}

First, we assumed that the accretion rate fluctuations at individual poles have similar variability properties as the total observed flux, which traces the total accretion rate. These are characterised by a typical red-noise-type power spectral density spectrum (PSD) with a break around the spin frequency. Therefore, we performed a series of simulations where for each instance two independent time series representing stochastic fluctuations of the accretion rate at two poles were simulated using the Python package \textit{pyLCSIM} \citep{Campana2017}. The broken power law parameters used for the simulations were chosen to be typical for accreting XRPs and are listed in \autoref{tab:simulated lc params} \citep{Fuerst2019, Moenkkoenen2022}. 
The observed contributions of the two poles as a function of phase define the mixing matrix in the simulation and were parameterised by a smoothed\footnote{For example, the \texttt{smooth} function from the SciPy cookbook can be used: \url{https://scipy-cookbook.readthedocs.io/items/SignalSmooth.html}} sum of six random Gaussian functions each to mimic typical XRP pulse profiles. An illustration of single-pole pulse profiles and their sum is presented in panel (a) of \autoref{fig:synthetic_dataset_pcva}. However, the Gaussian parameters were randomly varied, and therefore, this is just one of the realisations. Our simulations were based on 32 phase bins. The choice of phase bins depends on the source being studied and the required resolution of the pulse profile, and as PCVA is a statistical method it also depends on the count number, and therefore the brightness of the source, and the specific instrument used. A phase matrix containing the phase-resolved pulse-to-pulse light curves can be obtained by multiplying the simulated time series of accretion rate variations by the mixing matrix. Since variations in flux rather than accretion rate are observed in reality, we need to scale the simulated light curves and phase matrix to account for the limited counting statistics in real XRP observations. To achieve this, we use the obtained phase matrix to generate observed counts by drawing from a Poisson distribution defined by the average rate in the rate phase matrix. This step introduces random uncorrelated noise and reduces the overall degree of correlation, making the simulation more realistic.

We then use the NMF algorithm to decompose the simulated count rate phase matrices and estimate the mixing matrix to compare with the simulation input. A full example of the method and the detailed parameters used in this work is available as a Jupyter notebook\footnote{Available at DOI \url{https://doi.org/10.5281/zenodo.10369892}}. The recovered contributions are plotted as dashed lines in panel (a) of \autoref{fig:synthetic_dataset_pcva}. While the input shape is well reconstructed, there is a discrepancy in the relative scaling of the two poles. As mentioned above, this discrepancy is expected because BSS algorithms, including NMF, aim to find a decomposition of the observed signal into two maximally independent source signals, which may not correspond to the actual mixing if the signals are correlated to some degree. In the case of XRPs, the flux variability from the two poles is expected to be correlated at frequencies lower than the spin frequency, leading to a bias in any BSS reconstruction. In principle, one is not always interested in absolute scaling of the two components, that is, in many cases the relative scaling can also be hard to predict by theoretical models and in this case can be simply considered as additional parameter.

On the other hand, several approaches can be used to to recover the absolute scaling based on additional observables (i.e. phase-resolved spectroscopy or polarimetry), or, in principle, also using variability information. For example, the scaling can be adjusted by matching the simulated and observed correlation matrices -- constructed by calculating the degree of correlation between light curves at different phase intervals -- to reproduce the overall observed degree of correlation between the signals. In practice, this can be done by comparing the correlation matrices of the initial simulated phase matrix with a new correlation matrix based on a simulated light curve using the resulting NMF decomposition, which are shown in panels (b) and (c) in \autoref{fig:synthetic_dataset_pcva}. The decomposition can then be rescaled and a new phase matrix and associated correlation matrix computed. To determine the similarity between the two matrices, the sum of squared differences (SSD) or another similarity metric can be computed between the originally observed correlation matrix and the rescaled-decomposition-based correlation matrix. We used particle swarm optimisation for this, specifically the \textit{PySwarms}\footnote{\url{https://pyswarms.readthedocs.io/en/latest/}} module in Python, but alternative optimisers can also be used. Particle swarm optimisation is an algorithm in which a population of potential solutions, represented as particles, searches the parameter space to find optimal solutions. By mininmising the SSD, the optimal scaling can be obtained. The cost function used to perform the rescaling in Python is also included the example code\footnote{Available at DOI \url{https://doi.org/10.5281/zenodo.10369892}}.
The final rescaled components using this method are shown as dotted lines in panel (a) of \autoref{fig:synthetic_dataset_pcva}, and the correlation matrix based on the optimally scaled components is shown in panel (d). Note that in principle one could skip the NMF altogether and search for the mixing matrix directly comparing the correlation matrices, but this would require minimisation of a rather noisy cost function of 2x$n$ parameters, and is thus difficult to do via brute force minimisation. That is what NMF and other BSS methods are optimised for and do much better. On the other hand, we emphasise that phase dependence of the recovered components is not affected by this effect, and the optimisation only solves for two parameters (the two parameters used for the rescaling), which is much easier to achieve. 

We verified that this is indeed the case using a series of simulations as described above. In practice, the simulations were repeated 100 times and we found that this optimisation approach allows unambiguous recovery of the input data in all cases. To quantify this, the mean and standard deviation of the decompositions obtained within individual simulations can be calculated and compared with the input. This also provides a way of estimating the uncertainty for the recovered components associated with the recovery algorithm. An example is shown in panel (a) of \autoref{fig:synthetic_dataset_pcva}, which shows that the contribution from both poles is indeed robustly reconstructed with comparatively small uncertainties, which are shown in the upper right corner of the panel. The limitation of this approach is that in order to ensure that the derived scaling is correct, the simulated correlation matrices must capture the true variability properties of the data, which is not always easy to achieve. Otherwise, the reconstructed absolute scaling of the components may be biased, which ideally needs to be independently verified either by comparison with model predictions or by the use of independent observational data. We emphasise, however, that the shape of the reconstructed components appears to be insensitive to accurate simulation techniques, and the above caveat applies only to their relative scaling.

Applying the same procedure to simulated data corresponding to different count rates can also help to empirically estimate the minimum number of counts and pulse cycles required to robustly recover the simulation input. In particular, we ran a series of simulations tailored to mimic the observed \src count rate and found that a minimum of 1000 pulses and an average of 80 counts per pulse, or a minimum of 25 counts in each phase bin, are required to reliably recover components using PCVA, that is, the method appears to work well when each of the individual light curve phase bins is not dominated by Poissonian statistics. It should be noted that the data do not necessarily have to come from a single observation, as long as the shape of the pulse profile and thus the mixing operator remains unchanged. In this case, it can be assumed that the contributions of the two poles also remain constant and several observations could possibly be combined to meet the 1000 pulse requirement. We emphasise that the above estimates depend in principle on the actual shape of the two components, so simulations similar to those described above need to be carried out for any given source, using the initial decomposition as an input to the simulations to obtain a more reliable estimate. In any case, it is reasonable to expect that a robust reconstruction can only be obtained for reasonably high photon and pulse counts, that is, the method is mostly applicable to bright pulsars and data from instruments with large effective areas.

\section{Observations and PCVA analysis of Cen X$-$3}
\label{sec:application}
\src is a high-mass X-ray binary system that was first observed in 1967 by \citet{Chodil1967}, and the first X-ray pulsar ever discovered \citep{Giacconi1971}. The binary consists of a neutron star with a spin period of $\sim$4.8\,s orbiting an O-type supergiant companion star \citep{Schreier1972, Krzeminski1974}. It is a persistently accreting X-ray binary, although the observed flux is quite variable, which has been interpreted by \cite{2005A&A...442L..15P} as evidence for the existence of multiple accretion modes in the system. However, the origin of the observed variability is not entirely clear, as \cite{2011A&A...535A.102M} were unable to confirm the reported spectral changes. 

The pulse profile of \src is stable in time and consists of two peaks, the larger one being called the primary peak and the smaller one the secondary peak \citep{Raichur2010}. The pulse profile varies with energy, with the primary peak remaining almost unchanged between 1-40\,keV, while the secondary peak diminishes at higher energies (above $\sim$\,20\,keV). The pulse profile also appears to show some barely detectable variations with the orbital phase \citep{Raichur2010}, which are irrelevant for our study. 
The pulse profile decomposition analysis for the source was carried out by \cite{Kraus1996}, which we use as a baseline for the comparisons below. The results reported by \cite{Kraus1996} are not based on the same observations, but given the stability of the observed pulse profile shapes this is not expected to affect our conclusions.

\src has been extensively observed with several instruments. However, since high counting statistics are required for the application of the PCVA, we use here only observations of the source with the Rossi X-ray Timing Explorer (\textit{RXTE}) between Feb 28 and Mar 1, 1997 (MJD 50507-50508), obtained in the high accretion state of the source and covering a large part of the orbit. A total exposure time of 97\,ks was achieved during the observation with ID 20104-01-01-00.
All five proportional counters were switched on for most of the observation and the data were extracted for channels 10-35, corresponding to an energy range of 4-13\,keV and having a time resolution of $2^{-5}$\,s. This energy range was chosen because it contains most of the source photons, and there is no significant evolution of the pulse profile shape with energy within this band. The spin period of the pulsar during this observation was determined to be about $4.81426042$\,s \citep{klawin2023}. A visible dip associated with absorption in the companion's atmosphere was observed at the end of the observation. Therefore, only data before MJD 50508.38 were used for the analysis.
The corresponding phase matrix can be found at DOI 10.5281/zenodo.10149527\footnote{\url{https://doi.org/10.5281/zenodo.10149527}}.

We have applied the PCVA method to analyse the \textit{RXTE} data of Cen X-3 using the same procedures as for the simulated data, using 32 phase bins. The results, shown in panel (a) of \autoref{fig:results}, present the average decomposition and estimated uncertainties of the two single-pole pulse profiles. The primary peak of the total pulse profile is found to be a combination of two distinct single-pole components. Reflected and phase-aligned, both components are qualitatively similar. The secondary peak also appears to result from contributions from both components. The distance between the two maxima is $0.125$ ($45^\circ$) and between the two minima is $0.25$ ($90^\circ$). The minimum and maximum distance for each component is $0.1875$ in phase, corresponding to an angle of $67.5^\circ$. 

\begin{figure}[htbp]
	 \resizebox{\hsize}{!}{\includegraphics{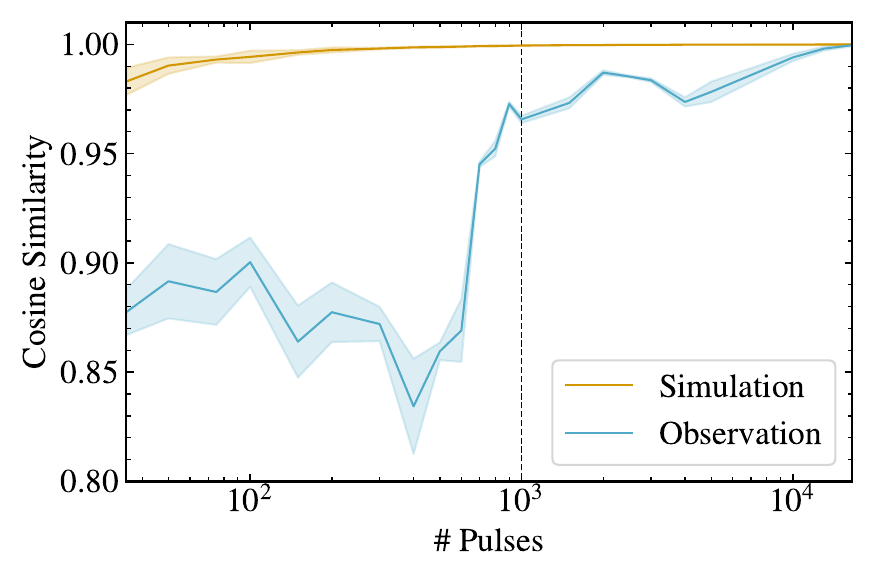}}
	 \caption{
  Minimum requirements for datasets suitable for PCVA analysis, determined from simulations and subsets of observed light curves. The figure shows the cosine similarity for PCVA results using the full light curve and subsets of varying length (expressed in number of pulses) for real (blue) and simulated (yellow) light curves. The mean and standard deviation of 10 PCVA results for each subset length are shown. In both cases, statistical limitations imply that the robustness of the reconstruction decreases as the quality of the data decreases (more rapidly for real data), allowing the minimum number of pulses required for analysis to be determined.
    }
	 \label{fig:npulses}
\end{figure}

\begin{figure}[htbp]
	 \resizebox{\hsize}{!}{\includegraphics{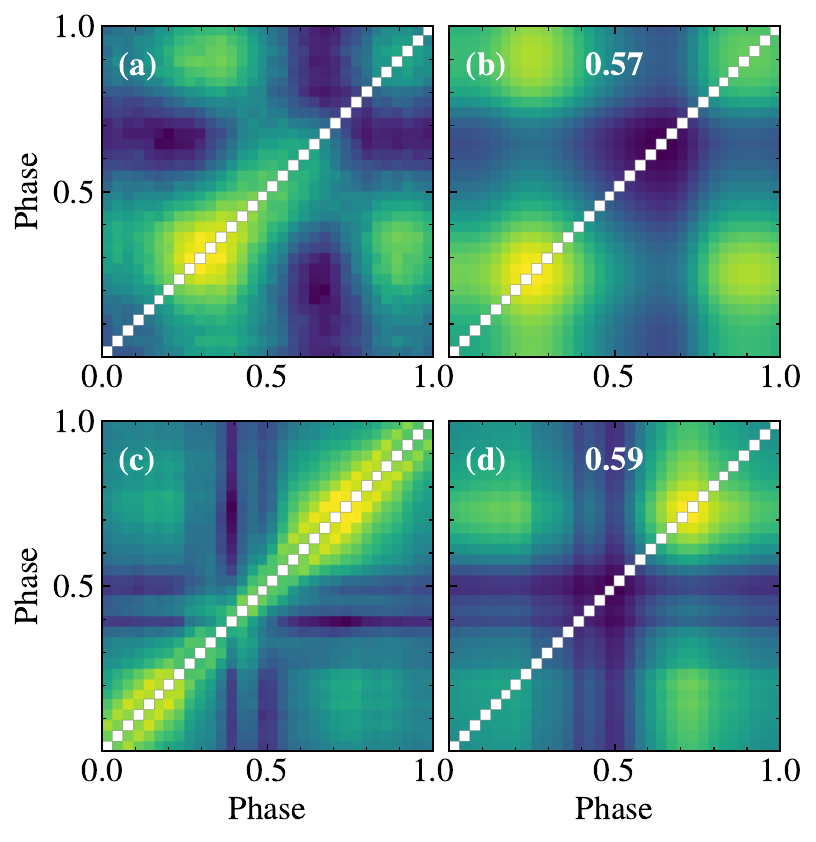}}
	 \caption{Comparison of correlation matrices derived from simulated and observed data. Panel (a) shows the correlation matrix of a simulation, while panel (b) shows the self-correlation matrix of the recovered components of PCVA for this simulation. The number in the panel indicates the Pearson correlation coefficient between panels (a) and (b). Similarly, panel (c) shows the correlation matrix of the observation and panel (d) shows the self-correlation matrix of the PCVA components together with the Pearson correlation coefficient between the two panels. Both simulated and observed data appear to recover the components similarly well. The colour scale is set individually for each panel between its minimum and maximum values to highlight the patterns.
     We note that panels (a) and (c) correspond to panels (b) in Figures \ref{fig:synthetic_dataset_pcva} and \ref{fig:results} respectively, while panels (b) and (d) differ from panels (d) in Figures \ref{fig:synthetic_dataset_pcva} and \ref{fig:results} and show the self-correlation matrix instead.
    }
	 \label{fig:fom}
\end{figure}

\begin{figure}[htbp]
	 \resizebox{\hsize}{!}{\includegraphics{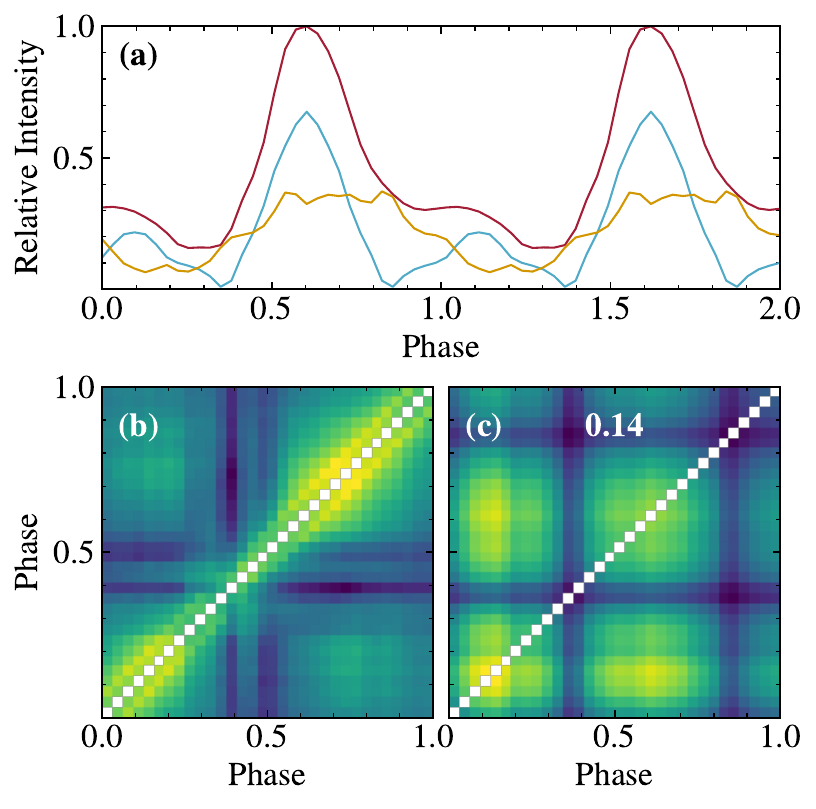}}
	 \caption{Comparison of correlation matrices from observations and the \citet{Kraus1996} components. Panel (a) shows the literature decomposition of \citet{Kraus1996} in blue and yellow, together with the pulse profile of \src in red. Panel (b) shows the correlation matrix of the observation and panel (c) shows the self-correlation matrix of the \citet{Kraus1996} components together with the Pearson correlation coefficient between panels (b) and (c). Note that the patterns seen in the correlation matrix are quite different and show additional features compared to those seen in the actual data in panel (b). The colour scale is set individually for each panel between its minimum and maximum values to highlight the patterns.
    }
	 \label{fig:kraus}
\end{figure}

By simulating a new phase matrix based on the PCVA decomposition, new correlation matrices can be constructed, as shown in panels (c) and (d) of \autoref{fig:results}, before and after re-scaling, respectively. We note the differences between the initially observed correlation matrix in panel (b) and the PCVA-based correlation matrix in panel (d). 
The observed discrepancy, which is not present in the same panels of the simulated data in \autoref{fig:synthetic_dataset_pcva}, leads us to question the applicability of PCVA to real data. In an ideal scenario, the performance of the method on real data would be assessed by comparing the decomposition result with the `ground-truth' single-pole pulse profiles. 
In the absence of this information, the method can be tested using simulations (where the inputs are known), but of course the validity of the results would depend to some extent on how closely the simulations resemble the real data. As a first step, however, we started with simulations using a cosine similarity as a metric to characterise the deviations of reconstructed pulses from the input, and found that this is always robust as long as the count statistics and the number of simulated pulse cycles are sufficient (where the actual numbers depend on both the source count rate and the shape of the input components). Another empirical test is to take subsets of the data of different lengths and look at where the reconstruction breaks down, that is, where it starts to show large variations between segments compared to the decomposition of the full length dataset, that is, to determine thresholds above which the decomposition becomes stable not only for simulated but also for real data.
This can be done both for the observed data and for more realistic simulations where the input shape of the components is assumed to be that recovered by PCVA rather than arbitrarily defined. The results are shown in \autoref{fig:npulses}. As can be seen from the figure, the quality of the reconstruction does indeed deteriorate as the number of pulses used for analysis decreases (for both real and simulated data, although the effect is weaker for simulations). However, above about 1000 pulses (dashed vertical line) the similarity levels off at over 96\% and remains consistently high above this threshold. The actual light curve used in our analysis contains 16500 pulses and is therefore sufficiently long for a robust decomposition.

The second test to assess the robustness of the PCVA result is to compare the actual correlation matrix with that obtained from simulations using the recovered components as input. This matrix encodes the variability information of the correlation between the two poles during the rotation of the neutron star. As can be seen in panels (b) and (d) of \autoref{fig:results}, the observed and simulated correlation matrices are not in perfect agreement, likely due to limitations of the simulations. 
Therefore, in order to assess the validity and provide a quantitative measure of the decomposition of the real data compared to the observed correlation matrix, we introduce a measure of correlation that includes only the information of the two decomposition components $\textbf{D1}$ and $\textbf{D2}$. To do this, we compute a `self-correlation' ($\textbf{SC}$) matrix:
$$\begin{gathered}
    \textbf{SC} = (\textbf{D1} \otimes \textbf{I} + \textbf{D2}^\intercal \otimes \textbf{I}) \cdot (\textbf{D1} \otimes \textbf{I} + \textbf{D2}^\intercal \otimes \textbf{I})^\intercal , \\
\end{gathered}$$
where $\otimes$ is the Kronecker symbol, $\textbf{I}$ is the (32,1) identity matrix, `$\cdot$' is the matrix multiplication and $\top$ is the transposition. 
The decomposition components, $\textbf{D1}$ and $\textbf{D2}$, correspond to the two column vectors in the mixing operator $B = [b_{ij}] \in \mathbb{R}^{n\times m=2}$ in terms of the classical BSS problem that was introduced in \autoref{sec:method}.
The terms in the brackets are linear combinations of the decomposition and thus combine the information of both, while the matrix multiplication gives a measure of the self-correlation of the combined decomposition. By computing this matrix, the decomposition can effectively be compared to the observed correlation matrix, and a correlation coefficient can be calculated between them to give a measure of similarity.
This self-correlation matrix was chosen because it allows only the inherent correlations of the decomposition (without simulated signals) to be calculated, thus providing a way of identifying patterns and structures.
In \autoref{fig:fom} the result is shown for both the simulated and the observed datasets. For the simulated data, we show the correlation matrix of an example simulation in panel (a) and the self-correlation matrix of the decomposition in panel (b). The average correlation coefficient between the two, out of 100 simulations, is $0.57$ with a standard deviation of $0.05$. The observed data clearly shows a similar pattern for the correlation matrix in panel (c) and the self-correlation of the decomposition in panel (d), and the Pearson correlation coefficient between the two is at a similar level to that of the simulations, at $0.59$. Thus, the method seems to work on the real data with a similar accuracy as the simulations. 
In addition, the shape and position of the features in the correlation matrix are noteworthy. As seen in panel (c), panel (d) of \autoref{fig:fom} also shows a `cross' shape, with an overall lower degree of correlation between the phases of 0.25 and 0.55. This suggests that the PCVA decomposition successfully accounts for the different contributions of the two poles during the rotation phase. The discrepancy between the correlation matrices shown in panels (b) and (d) of \autoref{fig:results} can therefore be attributed to the limited detail of the simulations, but does not seem to affect the decomposition and re-scaling capabilities of the method. This could be because we only simulate completely independent signals. In reality, however, the signals have some dependence, which reduces the overall correlation. It should be noted, however, that the correlation matrix is not used directly in the analysis. Instead, the phase-resolved light curves are used directly to recover the components. The degree of correlation is then only used to recover the scaling of the components. In principle, this step can be omitted, as the shape of the components can be used to model the pulse profiles, and the normalisation can be included as an additional parameter in the modelling. It is important to note that the NMF method itself is not affected by these considerations, as it is a purely statistical method that operates on real data (i.e. not fully independent data), with the only necessary assumption being that there exist some signals that are partially independent. We therefore take the result shown in panel (a) of \autoref{fig:results} as true.

\autoref{fig:kraus} panel (a) shows the decomposition obtained using the method proposed by \citet{Kraus1995, Kraus1996}. 
There are several notable differences between this decomposition and the PCVA composition shown in panel (a) of \autoref{fig:results}. First, the components recovered by PCVA are qualitatively similar, whereas the \citet{Kraus1996} method yields two significantly different components.
However, the most significant difference is that the PCVA decomposition results in components that are asymmetric in phase. This result directly contradicts the main assumption underlying the decomposition method proposed by \citet{Kraus1995}, which assumes that the emission characteristic at the two poles is symmetrical. As before, we compute the self-correlation matrix, see \autoref{fig:kraus}, and find that with this decomposition the correlation coefficient between the observed matrix, shown in panel (b) and the self-correlation matrix in panel (c) is much lower at $0.14$, compared to the $0.59$ of the PCVA decomposition shown in \autoref{fig:fom}. Furthermore, as can be seen in the figure, the observed pattern in the correlation matrix is quite different from the observed pattern. This is a very simple test, which in fact does not depend on any of the PCVA assumptions, and the observed discrepancy casts serious doubt on the validity of the assumptions of the \citet{Kraus1995} decomposition method.

To ensure the robustness of our results, we performed several tests. First, since \textit{NMF} is an iterative method that is not guaranteed to converge, we checked whether our conclusion could be invalidated by adjusting the convergence criteria or by changing the starting point for the optimisation. Specifically, we tried initialising the method using the original decomposition obtained by \citet{Kraus1996} and a symmetric version of our result as the starting point, rather than random values. We also tried initialising the method with two broad peaks corresponding to the peaks in the pulse profile, using the decomposition obtained by \cite{Kraus1996} as a starting point, and performing the analysis separately for several segments within the observation. In all cases, the original input was recovered and the method converged back to the expected solution shown in panel (a) of \autoref{fig:results}. 
We stress that the fact that PCVA converges to the same result regardless of the details of the analysis, and that the result differs from the \cite{Kraus1996} decomposition, also implies that the latter is inconsistent with the observed phase-phase correlation properties. This can be illustrated by inspecting the self-correlation matrix for the \cite{Kraus1996} decomposition presented in panel (c) of \autoref{fig:kraus}. It is clear that the self-correlation matrix appears to be quite different from the observed one, that is, different parts of the pulse profile are expected to change together compared to what is actually observed. We therefore conclude that the obtained decomposition is robust and the discrepancy with the \citet{Kraus1996} result is real.

\begin{figure*}[htbp]
	 \includegraphics[width=\textwidth]{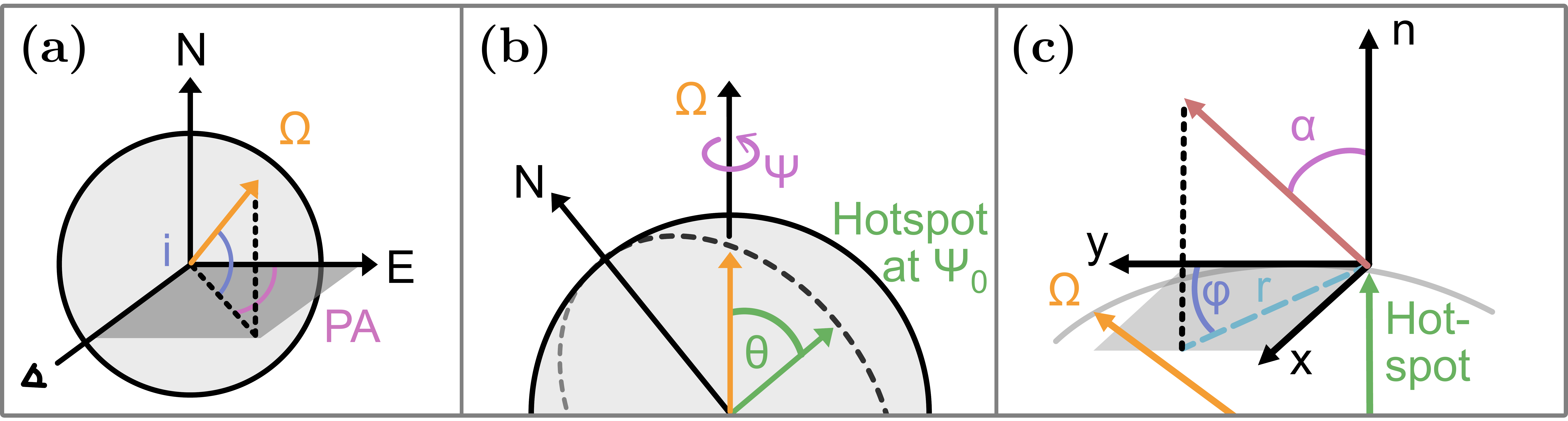}
    \centering
	 \caption{Pulsar geometry and the beam pattern of the toy model. In panel (a) the pulsar geometry is shown as a sketch with the spin axis ($\Omega$) located by the inclination ($i$) and the position angle of the pulsar spin (PA). The position of the emission region on the surface with respect to the spin axis is determined by the magnetic co-latitude $\theta$ and the rotation phase is denoted by $\Psi$, shown in panel (b). Panel (c) shows that the beam pattern is defined by the angles $\alpha$ and $\varphi$, where $\alpha$ is measured from the normal and the reference axis for $\varphi$ is the $y$ axis, which points towards the spin axis which corresponds to north when the emission region is at $\Psi_0$.
  }
	 \label{fig:polar}
\end{figure*}

\section{Discussion}
The main conclusions about the emission from each neutron star pole that can be drawn from the analysis described above are that a) the observed emission pattern from each of the two poles is intrinsically asymmetric in phase, and b) the contributions from the two poles are rather similar, although reflected in phase. 
To better understand the reasons for these observations, we performed a simplified modelling of the recovered components, aiming to recover the intrinsic beam pattern local to the emission region location. 

The most basic model involves two areas of emission, confined to the surface of the neutron star, with no extended accretion columns above the surface. However, this is likely not entirely accurate for \src. It is believed that for this source, the emission originates from a fan plus pencil beam \citep{Kraus1996, 2022ApJ...941L..14T}. The findings of \citet{2022MNRAS.517.4138B} align with this conclusion. Their study analysed pulse profiles of \src in varying luminosity states and discovered substantial changes, which may be attributed to the onset of an accretion column at higher luminosities. As a result, the accretion column should play a role in the pulse profile. However, it should be noted that the primary peak of the pulse profile remains at the same phase (see Fig. 3 in \citealt{2022MNRAS.517.4138B}), and therefore the emission should still be dominated by the pencil beam, with some additional contributions from a fan beam. Therefore, modelling the hot spots is a reasonable first approximation. This is a limitation of the current model; however, as it is intended for illustration purposes only, we have chosen the most straightforward model feasible without overly complicating the beam pattern.

As discussed by \cite{Kraus1995}, the fact that the observed phase curve produced by a single pole is asymmetric in phase implies that the intrinsic beam pattern local to the emission region must also be asymmetric. Such a situation can occur, for example, if the magnetic field (which defines the opacities) is not orthogonal to the photosphere of an emission region on the surface, but is slightly tilted. In this case, the local beam pattern may be symmetric about some direction, but asymmetric with respect to the normal, which, with rotation of the neutron star, can produce asymmetric pulses.

To model such a situation, we define the geometry of a pulsar as follows. First, the orientation of the neutron star's spin axis and the location of the emission region on the surface must be defined. Panel (a) of \autoref{fig:polar} shows the location of the spin axis, which is defined by the inclination ($i$) and position angle (PA) of the pulsar's spin ($\Omega$). In panel (b), the location of the emission region is defined by the magnetic co-latitude ($\theta$) and the phase of the rotation around the spin axis ($\Psi$). The emission in our toy model is produced by a single emission region at each pole. To define the locally asymmetric emission pattern we need to define two additional angles, $\alpha$ and $\varphi$, shown in panel (c) of \autoref{fig:polar}.
We can express these angles in terms of Cartesian coordinates represented by $x$ and $y$, as shown in panel (c). Note that in the unit sphere, the projection of the angle $\alpha$ onto the $x$-$y$ plane is equal to the distance $r$ from the origin. Then we have
$$\begin{gathered} 
    \alpha = \arcsin{(r)}, \text{ with } r = \sqrt{x^2 + y^2} \\ 
    \varphi = \arctan\left(\frac{y}{x}\right) \text{ .} 
\end{gathered}$$

To parameterise the local beam pattern we use the $x$-$y$ coordinates and a two-dimensional cosine:
$$\begin{gathered}
    f(\delta x_1, \delta y_1, \gamma) = 
    \left[\cos(\delta x_1 ) \cos(\delta y_1)\right]^\gamma \text{ ,}
\end{gathered}$$

where, $\delta x_1$ and $\delta y_1$ shift the cosine in the $x$-$y$ plane, shown in \autoref{fig:polar} panel (c), and can each take values between $-\pi$ and $\pi$. 
The function can be made narrower or `peaky' by raising it to the power of $\gamma$.
We also account for gravitational light bending in our model using the analytical approximation of \citet{Beloborodov2002}:
$$\begin{gathered}
    1-\cos \alpha_\text{em} = (1-\cos \psi)\left(1-\frac{r_g}{R}\right) \text{ ,}
\end{gathered}$$

where $\alpha_\text{em}$ is the angle at which a photon is emitted with respect to the radius and $\psi$ is the angle at which an observer sees the emission of the photon. We use the typical value of the radius of a neutron star $R \approx 3 r_g$, where $r_g$ is the Schwarzschild radius. 

We emphasise that the above parametrisation is only a simple phenomenological function to describe a situation where the local emission pattern is asymmetric, reflecting the fact that the actual beam pattern is unknown and most likely not symmetric. The toy model we construct is therefore only intended to illustrate that it is possible to describe asymmetric pole contributions obtained from PCVA (and the overall pulse profile) even if the assumptions regarding the actual beam pattern are only minimally relaxed.

\begin{figure*}[htbp]
	 \resizebox{\hsize}{!}{\includegraphics{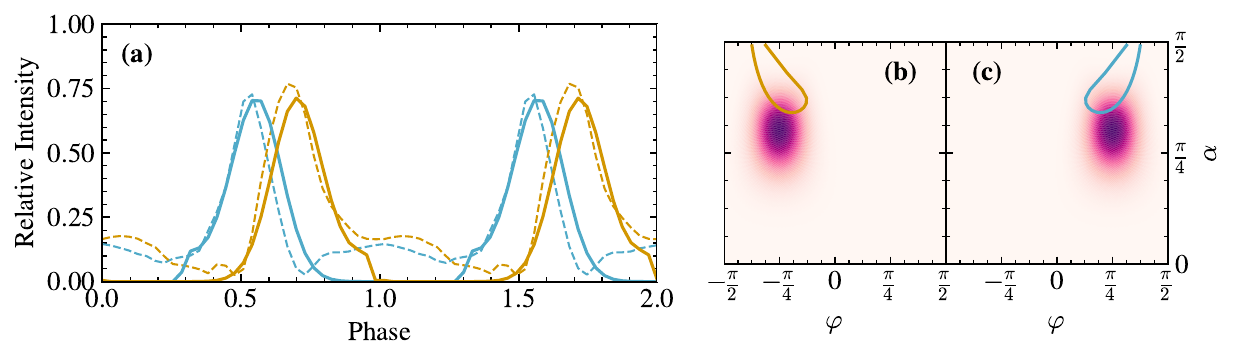}}
	 \caption{Example of toy model results. Panel (a) shows the single-pole pulse profiles generated with a toy model using the system parameters described in the text. The PCVA components are shown as dashed lines for reference.
  Panels (b) and (c) schematically show the beam patterns of the two poles. The relative intensity is colour coded between 0 (white) and 1 (dark purple). The lines indicate the parts of the beam patterns that can be seen by a distant observer and the colour correspond to the pulse profiles of panel (a). The simple model allows for a displacement of the dipole configuration and the symmetry axis of the beam patterns is allowed to be offset from the normal and different for each emission region. In this example, this results in asymmetric pulse profiles that appear reflected in phase.}
	 \label{fig:toy}
\end{figure*}

The basic geometry of \src was recently determined by \citet{2022ApJ...941L..14T}, who used an orbital inclination of about $70^\circ$ to obtain the position angle of the pulsar spin of about $49^\circ$ and a magnetic co-latitude of about $16^\circ$. We investigated which intrinsic beam pattern would correspond to the single-pole pulse profiles derived from the PCVA method analysis using the parameters listed in \autoref{tab:toy model}, but notably changing the position angle from the \citet{2022ApJ...941L..14T} geometry, as discussed below.

\begin{table}
\caption{Parameters of the toy model. Hotspots 1 and 2 correspond to the beam pattern in panels (b) and (c) of \autoref{fig:toy}, respectively. For a definition of the parameters, see \autoref{fig:polar}.}
\label{tab:toy model}
\centering
\begin{tabular}{lcc}
	\hline\hline
	Parameter & Hotspot 1 & Hotspot 2 \\ 
	\hline
	$i$ & 70$^\circ$ & $-70^\circ$ \\
    PA & 90$^\circ$ & $-90^\circ$ \\
    $\theta$ & 16$^\circ$ & 16$^\circ$ \\
    $\delta x_1$ & 0.7 & 0.7 \\
    $\delta y_1$ & $-0.7$ & 0.7 \\
    $\gamma$ & 20 & 20 \\
    $\Psi$ & 1.4 & 0 \\
	\hline
\end{tabular}	
\end{table}

An example is shown in \autoref{fig:toy}. Panels (b) and (c) show the beam patterns of the two emission regions with the parameters listed in \autoref{tab:toy model}. With the given geometry, only the part of the beam pattern traced by the blue and yellow lines can be seen by a distant observer. The resulting pulse profiles are shown in their respective colours in panel (a), along with the PCVA decomposition as dashed lines for reference. Note that we have selected identical beam patterns for both poles, only mirrored in direction $\varphi$, with their most intensity maxima at approximately $-\pi/4$ for panel (b) and $\pi/4$ for panel (c).
This example illustrates that an asymmetric and mirrored beam pattern results in asymmetric pulse profiles, which are reflected in phase. In order to roughly align the maxima with the PCVA result, a relaxation of the strict dipole with a phase offset is required in this model. However, the non-zero PCVA result implies that both poles are visible for a significant portion of the rotation, even with the influence of gravitational light bending, and have comparable overall visibility. To achieve this outcome in the present model, the position angle must be kept near 90 degrees, but a considerably smaller magnetic co-latitude would be necessary. An alternative viable approach could involve a more complicated model that integrates the flux over extended emission regions, such as arcs on the surface of the neutron star. This would generate a more accurate representation of the physical system, and studies by \citet{Postnov2013} and \textit{NICER} results by \citet{Riley2019} suggest that the emission region of X-ray pulsars may be more complicated than previously thought. Incorporating this possibility in the model would be more realistic and could potentially solve the problem of the visibility of the emission region in the toy model. An extended emission region would allow fragments of the emission region to remain visible for longer periods of time during the rotation period. 
In conclusion, despite the lack of complexity of the toy model compared to the real system, it shows that certain key aspects of the PCVA decomposition can be explained by a relatively simple geometry and intrinsic beam pattern, provided that certain asymmetries are present.
The recovered contributions can now be used in the future as a basis for more advanced models, which are beyond the scope of this paper.

\section{Summary}\label{sec:summary}
The results can be summarised as follows:
\begin{enumerate}
    \item Statistical fluctuations affect the accretion rate to the two poles of XRPs, and we can exploit the lack of full correlation between the accretion at each pole to disentangle the contributions to the pulse profile using blind source separation.
	\item Given the properties of our data, non-negative matrix factorisation (NMF) was used for the blind source separation.
	\item We developed and verified the phase correlated variability analysis (PCVA) technique using simulated data. The simulated light curves were based on a broken power law spectral model and known single-pole pulse profiles. PCVA successfully recovered the initial pulse profiles of the two poles and optimisation was performed using correlation properties. 
    \item The new PCVA method was applied to \textit{RXTE} data from \src. This resulted in two distinct single-pole pulse profiles of approximately equal amplitude and width. The profiles appear asymmetric and reflected in phase.
	\item Our results are not in agreement with \citet{Kraus1995, Kraus1996} since the profiles are not intrinsically symmetric and do not fulfil the assumptions of their method.
	\item We have used a toy model with defined geometry and an emission region with a simple asymmetric beam pattern to illustrate the capabilities of the PCVA to reliably and accurately recover pulse profiles.
 \end{enumerate}

In conclusion, the PCVA method can be applied to any bright XRP, and we provide a recipe for future studies. A fully working example of the method on simulated data is made available to the wider scientific community as a Jupyter notebook\footnote{The code accompanying this paper is available as version 1.1.0 under the following link: \url{https://doi.org/10.5281/zenodo.10369892}. Please note that the latest version of the code can be found at \url{https://doi.org/10.5281/zenodo.10159881} for the most recent features and bug fixes.}, in line with the principles of reproducible research to advance understanding in the field.
The results of the analysis presented in this paper provide an alternative perspective on the emission regions of XRPs and can be seen as complementary to the polarimetric studies.
Observations with \textit{IXPE} and the upcoming \textit{eXTP} observatory could provide ideal data for future studies as both our method and polarimetry require high counting statistics and thus both analyses can be done using the same data-set.

\begin{acknowledgements}
  The authors thank the anonymous reviewer for useful comments and suggestions that improved this manuscript.
  IS, VD would also like to thank the Deutsche Forschungsgemeinschaft (DFG) for the support (DO 2307/1-1).
\end{acknowledgements}


\bibliographystyle{aa/aa}
\bibliography{bibliography}

\end{document}